\documentclass[a4paper, oneside, reqno, 11pt]{article}
 
\usepackage[a4paper, top=1.6in, bottom=20ex, left=35mm,
right=34mm,headheight=28pt,headsep=8pt]{geometry}
\usepackage{graphicx}
\usepackage{epsfig}
\usepackage{amsthm}
\usepackage{amsmath}
\usepackage{latexsym}
\usepackage{amsfonts}
\usepackage{amssymb}
\usepackage{hyperref} 
\usepackage{float}       

\begin{document}

 \title{\LARGE \textbf{Spatial dispersion of elastic waves in a bar characterized by tempered nonlocal elasticity}}
 \author{{Vikash Pandey\footnote{Corresponding author. Electronic address: vikashp@ifi.uio.no} $^1$, 
Sven Peter N\"{a}sholm$^2$, Sverre Holm$^3$}\\ 
$^{1,3}$\footnotesize   \emph{Department of Informatics, University of Oslo, P.O. Box 1080, NO-0316 Oslo, Norway}\\
$^{2}$\footnotesize   \emph{NORSAR, P.O. Box 53, N-2027 Kjeller,  Norway}\\}

\date{April 28, 2016}
\maketitle

 \begin{abstract}

We apply the framework of tempered fractional calculus to investigate the spatial dispersion of elastic waves in a one-dimensional elastic bar characterized by range-dependent nonlocal interactions. The measure of the interaction is given by the attenuation kernel present in the constitutive stress-strain relation of the bar, which follows from the Kr\"{o}ner-Eringen's model of nonlocal elasticity. We employ a fractional power-law attenuation kernel and spatially temper it, to make the model physically valid and mathematically consistent. The spatial dispersion relation is derived, but it turns out to be difficult to solve, both analytically and numerically. Consequently, we use numerical techniques to extract the real and imaginary parts of the complex wavenumber for a wide range of frequency values. From the dispersion plots, it is found that the phase velocity dispersion of elastic waves in the tempered nonlocal elastic bar is similar to that from the time-fractional Zener model. Further, we also examine the unusual attenuation pattern obtained for the elastic wave propagation in the bar.

 \medskip


 \smallskip


 \smallskip

\texttt{The peer-reviewed version of this paper is published in\\
\textit{Fract. Calc. Appl. Anal. Vol. 19, No 2 (2016), pp. 498-515,\\\href{DOI: 10.1515/fca-2016-0026}{DOI: 10.1515/fca-2016-0026}}, and is available online at \\
\textit{\href{http://www.degruyter.com/view/j/fca}{http://www.degruyter.com/view/j/fca}}\\
The current document is an e-print which differs in e.g.  pagination, reference
numbering, and typographic detail.}

 \end{abstract}

 \maketitle

 \vspace*{-16pt}


\section{Introduction}\label{sec:1}

\setcounter{section}{1}
\setcounter{equation}{0}

The frequency dependency of the phase velocity and wave attenuation originating from the wave-medium interaction is mathematically expressed by the dispersion relation $k\left(\omega\right)$, where $k$ and $\omega$ are the wavenumber and angular frequency of the wave respectively. Since it is possible to extract the viscoelastic parameters of a material from its dispersion curve, the investigation to obtain an accurate dispersion relation becomes pivotal for many applications in the field of acoustics, medical imaging, seismology and geophysics \cite{Meidav1964, Wells1975, Casula1992, Szabo1995, Klatt2007, Zhang2008, Mueller2010}.
\par

One of the methods which is followed to model dispersion assumes that the material possesses memory, which can be classified into two types; memory in time and memory in space \cite{Agarwal1974, Eringen1992, Eringen2002, Holm2013}. In materials with time nonlocality or simply memory, the effect at a given point in space at a given time is dependent on the causes at the same point for all preceding times. On the other hand, in materials with space nonlocality or spatial memory, the effect at a given point in space at a given time is dependent on the causes at all points in space. In this paper, we differentiate the dispersion arising due to time nonlocality and spatial nonlocality by referring to them as "temporal dispersion" and "spatial dispersion" respectively.
\par
For an elastic material with temporal memory, physically it implies that the material remembers its past deformations. Such materials do not obey the simple Hookean elasticity, but rather they fall into the category of viscoelasticity. The classical viscoelastic models, e.g. the Maxwell, Kelvin-Voigt and Zener constitutive stress-strain models, lead to temporal dispersion and attenuation of waves in materials \cite{Meidav1964, Casula1992, Mainardi2010}. The basic building blocks of the models are elastic springs and viscous dashpots, which are arranged in series and (or) parallel combinations to depict the interplay between the elastic and viscous properties of a material. The fractional order counterpart of the classical viscoelastic models which are studied using fractional calculus, have been found even more useful in investigating the dispersive properties of a wide range of complex materials (e.g. biological tissues, polymers and earth sediments) \cite{Zhang2008, Naesholm2011, Holm2014, Holm2013, NaesHolm2013a, NaesHolm2013b, Straka2013}.
\par
For the last fifty years \cite{Kroner1967, Eringen1972a, Eringen2002}, it has been well established that under the action of a deforming stress, the constituent points of an elastic material display range-dependent nonlocal interaction between themselves. Consequently, the applied stress is not confined to a local point, but rather distributed to all the interacting points of the material. A wave travelling in such a nonlocal material undergoes spatial dispersion, in addition to the regular temporal dispersion \cite{Eringen1972b}. It is essential to emphasize that temporal dispersion manifests itself as a result of the localized wave-material interaction which is primarily dependent on the frequency $\omega$. But since space nonlocality in a material is dictated by the wavelength of the traversing wave, the resulting spatial dispersion is determined by the wavenumber $k$.
\par
Lately, spatial nonlocal operators of the form of fractional Laplacians have been used to model dispersive properties of acoustic media \cite{Carcione2010, Zhu2014}. However, the study of spatial dispersion has not received as much attention as the temporal dispersion \cite{Eringen1992, Hanyga2012}. The main factor behind this is the satisfactory explanation of dispersive properties of most materials by temporal dispersion, which is modelled under the framework of classical theory of elasticity. However since the classical theory is based on the principles of continuum mechanics, it assumes localized stress-strain in the material, and is therefore not applicable in problems where nonlocal interactions play a dominant role \cite{Eringen2002}.
\par
The primary motivation for our work is that much of the current research (see, e.g. \cite{Paola2008, Carpinteri2011, Hanyga2012, Sapora2013, Paola2013, Tarasov2013}) has focused more on the mechanical properties of nonlocal elastic materials than its dispersive properties. Second, the results from such studies may even influence how new materials can be artificially engineered \cite{Banerjee2011}. Besides, we will also illustrate how the framework of tempered fractional calculus can be used to overcome the boundary value problems which are often encountered when nonlocal elasticity is modelled using fractional calculus. The purpose of this paper is therefore to modify the nonlocal elasticity in an infinitely long one-dimensional bar with tempering, and then investigate the spatial dispersion of elastic waves in the bar. Further, we show how different numerical techniques can be used to solve the spatial dispersion relation which is often analytically difficult.
\par
The rest of the paper is organized as follows. In Section~\ref{sec:2}, the suitability of fractional calculus to study nonlocal elasticity problems is analysed in the light of some recent studies. In Section~\ref{sec:3}, we spatially temper the conventional power-law attenuation kernel of the nonlocal elastic bar and examine its significance. Then in Section~\ref{sec:4}, we utilize the mathematical framework of tempered fractional calculus to derive the spatial dispersion relation from the constitutive equation of the bar. In Section~\ref{sec:5}, we employ numerical techniques to solve the dispersion equation for complex wavenumbers $k$ and subsequently obtain the dispersion plots. Further, we compare temporal dispersion and spatial dispersion by pointing out the resemblance of the phase velocity dispersion curve obtained from the space-fractional tempered nonlocal elastic model with that from the time-fractional Zener model. Moreover, the spatial dispersion relation anticipates an unusual attenuation behaviour for the elastic wave propagation in tempered nonlocal elastic bar, which we justify by physical arguments. Finally in Section~\ref{sec:6}, the potential implications of this work are used to draw some conclusions.

\section{Nonlocal elasticity and the framework of fractional calculus}\label{sec:2}

\setcounter{section}{2}
\setcounter{equation}{0}

Nonlocal continuum field theories describe the physics of materials whose behaviour at a given local point is determined by the state of all points of the body (see, \cite{Kroner1967, Eringen1972a, Eringen1972b, Eringen2002}). The theory is built upon two assumptions. First, the molecular interactions in a material are inherently nonlocal. Second, the theory assumes the energy balance law to be valid globally for the entire body. The comprehensive and robust formulation of nonlocal theory is ascertained by its wide range of applications \cite{Eringen1992}. Lately, nonlocal theory has also found applications in nanomechanics which integrates solid mechanics with atomistic simulations to analyse the strength and dispersive properties of carbon nanotubes (CNTs) \cite{Wang2005}, \cite{Sundararaghavan2011}.
\par
According to the Kr\"{o}ner-Eringen's model of nonlocal elasticity, the total strain response of a material to a deforming stress $\sigma\left(x\right)$ is the sum of the local strain $\epsilon\left(x\right)$ and, the nonlocal strain \cite{Kroner1967}, \cite{Eringen1972a}. The nonlocal strain is given by the convolution integral of the strain in the neighbourhood $\epsilon\left(y\right)$ with the attenuation kernel $g\left(x-y\right)$, where $y$ is the Euclidean distance from the local point $x$. For an infinitely long, one-dimensional, isotropic, linear nonlocal elastic bar, the constitutive equation is then given as
\begin{equation}\label{eq1}
\sigma\left(x\right)=E_{0}\left[\underbrace{\epsilon\left(x\right)}_{\text{Local strain}}+\underbrace{\kappa\int\limits _{-\infty}^{\infty}\epsilon\left(y\right)g\left(x-y\right)dy}_{\text{Nonlocal strain}}\right],
\end{equation}
where $E_{0}$ is the elastic modulus at zero frequency. The material constant $\kappa$ represents the strength of nonlocality in the material. As seen, in the limit as $\kappa\rightarrow0$, equation \eqref{eq1} reduces to the case of Hooke's law. It is reasonable to assume the bar to be one-dimensional provided the bar thickness on both sides is much smaller than the wavelength of the propagating wave. Besides, an infinite length of the bar ensures complete attenuation of the wave by the time it reaches its free ends, which makes the wave propagation free from reflections and the resulting complexities of boundary value problems. An illustration of the equivalent mechanical model of nonlocal elasticity can be found in \cite{Paola2008, Carpinteri2011, Paola2013}, where points of a material are shown connected to each other by means of springs of different stiffnesses.
\par
Although some suitable forms of the attenuation kernel have been suggested in the relevant literature (see, e.g. \cite{Polizzotto2001, Eringen2002, Sundararaghavan2011, Zingales2011, Paola2013}), recently there has been a growing interest in the power-law kernel (see, \cite{Paola2008, Atanackovic2009, Carpinteri2011, Sapora2013, Challamel2013}), which is given as
\begin{equation}\label{eq2}
g\left(y\right)=\frac{1}{\Gamma\left(1-\alpha\right)\left|y\right|^{\alpha}},
\end{equation}
where $\alpha\in\left[0,1\right]$ is the fractional exponent and $\Gamma\left(z\right)$ is the Euler's Gamma function defined for a complex variable $z$ as:
\begin{equation}\label{eq3}
\Gamma\left(z\right)=\int\limits _{0}^{\infty}x^{z-1}e^{-x}dx, \text{ } \Re\left(z\right)>0.
\end{equation}
The advantages of opting for a fractional power-law attenuation kernel are three fold. First, the exponent $\alpha$ allows a wide range of distance decaying interactions to be taken into account and it also plays the role of a scale parameter which is inherent to the theory of nonlocal elasticity \cite{Paola2008}. Second, the similarity between the attenuation kernel and the power-law memory kernel of fractional calculus allows the mapping of nonlocal problem under the comprehensive framework of fractional calculus. Third, power-law solutions are often encountered in experimental measurements which further justifies fractional modelling of the nonlocal elasticity.
\par
Here, it is essential to clarify that the attenuation kernel $g\left(y\right)$ should not be confused with the term "memory kernel" which is often used by the acoustic community for the temporal power-law functions.				
\par
Following the commutativity property of convolution and also considering the fact that the attenuation kernel is symmetric across the local point of an isotropic bar, we substitute equation \eqref{eq2} in equation \eqref{eq1} and obtain the constitutive relation as,
\begin{equation}\label{eq4}
\sigma\left(x\right)=E_{0}\left[{\epsilon\left(x\right)}+{2\kappa\int\limits _{0}^{\infty}\frac{\epsilon\left(x-y\right)}{\Gamma\left(1-\alpha\right)y^{\alpha}}dy}\right].
\end{equation}
Clearly, the absence of a Dirac-delta form of the attenuation kernel followed by the presence of an integral in equation \eqref{eq4} enforces strong nonlocal elasticity in the material. In addition, it is seen that the integral term in equation \eqref{eq4} is similar to a one-dimensional fractional Laplacian of order $\left(1-\alpha\right)$, which also implies that the attenuation kernel given by equation \eqref{eq2} has indeed the form of a Riesz kernel \cite{Samko1993, Chen2004}.
\par
Although the framework of fractional calculus offers a powerful tool for investigating problems characterized by fractional power-law kernels, it suffers with an inherent problem of singularity which has also been recently reported by one of its proponents in \cite{Caputo2015}. In the limit as ${y\rightarrow 0}$ in equation \eqref{eq2}, $g\left(y\right)\rightarrow \infty$; this implies that the nonlocal contribution to the stress at the local point $x$ is infinite, consequently the displacement at the local point $u\left(0\right)$ also grows to infinity, which is clearly non-physical. This is evident mathematically as well, since the nonlocal integral in equation \eqref{eq4} cannot pass the p-test of convergence and therefore never converges to a real value (see, Chapter 1 in \cite{McQuarrie2003}).
\par
Most works on the topic have encountered the same problem. A recent work (see, \cite{Paola2008}) which investigates the strength of a nonlocal elastic bar of finite length $L$ characterized by the same kernel as given by \eqref{eq2}, considers the divergent nature of the fractional integral as an open problem. The authors further trace this to the original work by Kr\"{o}ner and suggest enforcing the condition: $u\left(0\right)=u\left(L\right)=0$, to circumvent the boundary value problem \cite{Kroner1967}. Further, a very recent article (see, \cite{Challamel2013}) overcomes one of the boundary value problems by assuming the nonlocal bar to be of infinite length, however it overlooks the divergent behaviour of the nonlocal integral at the local point. Another paper on the topic (see, \cite{Carpinteri2011}) transforms the problem to the case of $1 < \alpha < 2$ by expressing the Riesz fractional derivatives in the Marchaud-like form. An entirely different solution is suggested in \cite{Polizzotto2001}, where the author considers the distance $y$ to be geodetical, instead of the usual Euclidean. It is worthwhile to mention that the article \cite{Atanackovic2009} (see, Lemma 1) suggests the possibility of a solution of the wave equation for a nonlocal elastic bar in a tempered space distribution which does not requires enforcing of boundary conditions.
\par
To summarise, we find that for the case $0 \leq \alpha \leq 1$, if homogeneous boundary conditions are not imposed, the nonlocal elastic model characterised by the conventional power-law kernel becomes non-physical and mathematically inconsistent. Consequently, a closed form expression of the spatial dispersion relation has never been obtained for a nonlocal elastic material, and thus the frequency dependency of phase velocity and wave attenuation has remained unanalysed.

\section{Spatial tempering of the attenuation kernel}\label{sec:3}

\setcounter{section}{3}
\setcounter{equation}{0}

Now, we follow a similar approach as mentioned in \cite{Atanackovic2009} to overcome the boundary value problem associated with the fractional nonlocal integral. The method is inspired by the exponential truncation of arbitrary large flights produced by Levy stable distributions (see, \cite{Cartea2007, Meerschaert2011}). Accordingly, we temper the conventional fractional attenuation kernel \eqref{eq2} as
\begin{equation}\label{eq5}
g\left(y\right)=\frac{1-e^{-\lambda \left|y\right|}}{\Gamma\left(1-\alpha\right)\left|y\right|^{\alpha}}\,,
\end{equation}
where $\alpha\in\left[0,1\right]$ and $\lambda > 0$ is the tempering parameter.
\par
\begin{figure} 
\centering
\includegraphics[scale=0.3]{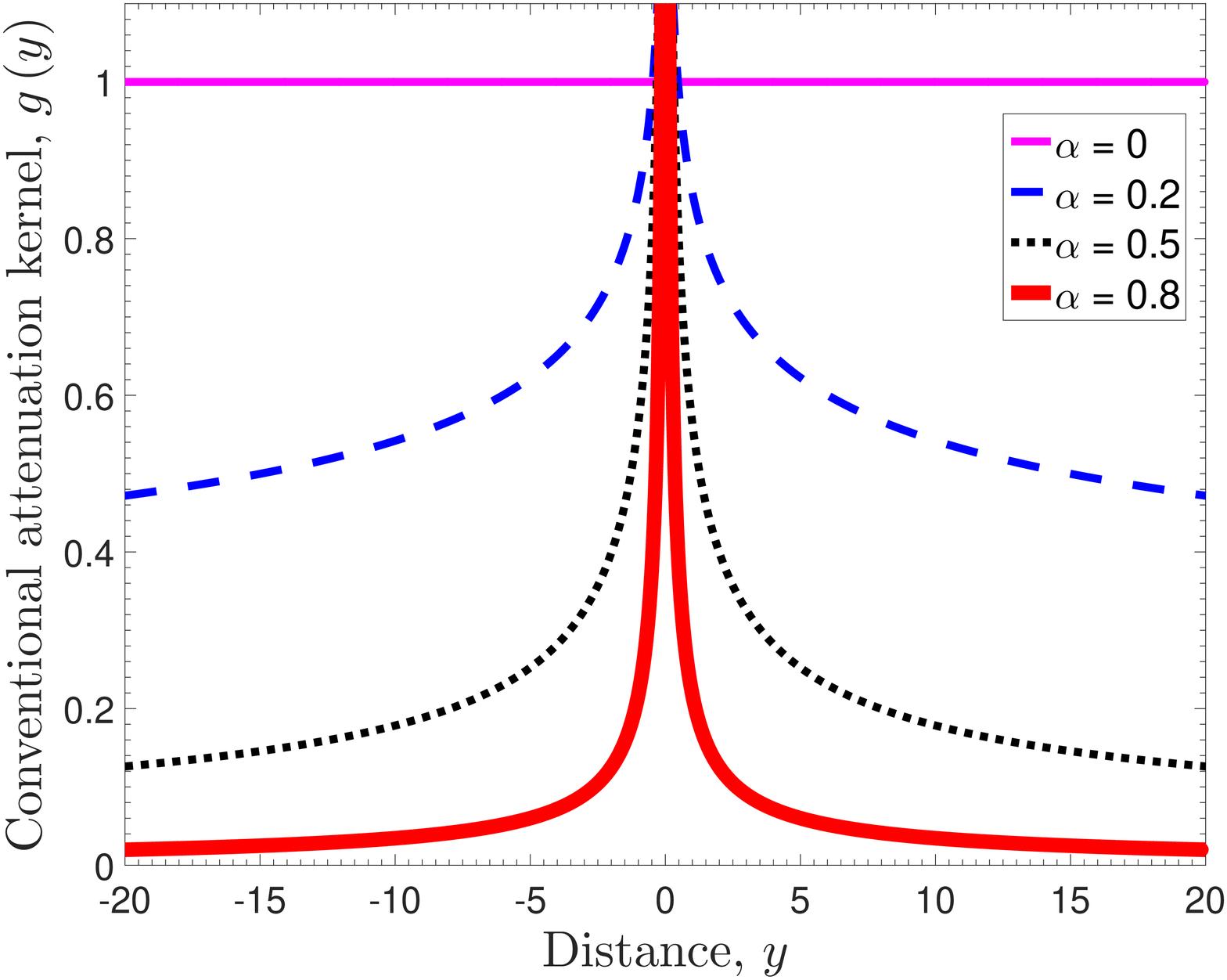}
\includegraphics[scale=0.3]{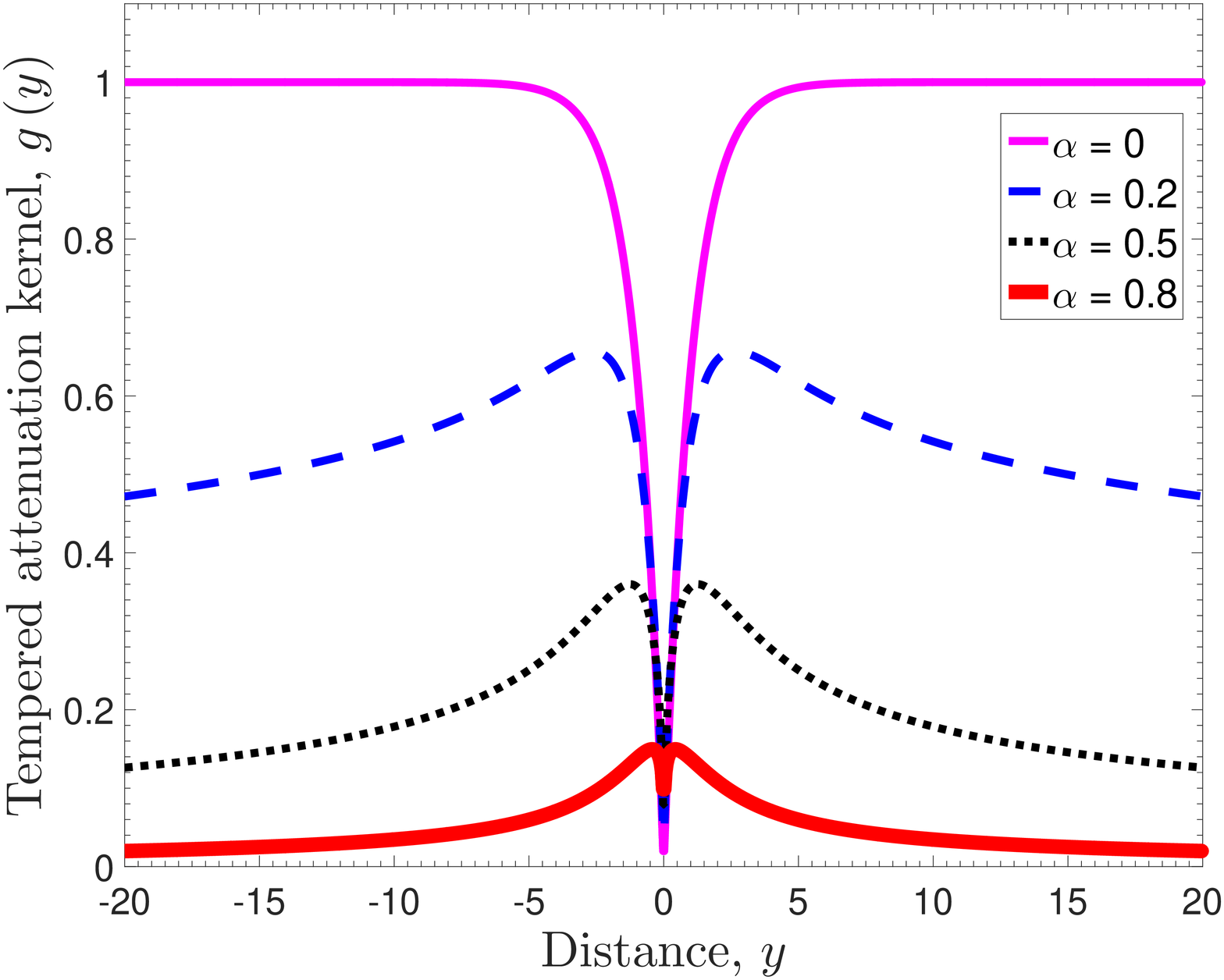}
\smallskip
\caption{\label{fig:Attenuation}Comparison of the conventional attenuation kernel (top pane) with the tempered attenuation kernel (bottom pane) for $\lambda=1$. Local point corresponds to $y=0$. The fractional order $\alpha$ has values $0$ (magenta, solid curve), $0.2$ (blue, dashed curve), $0.5$ (black, dotted curve) and $0.8$ (red, thicker solid curve).}
\end{figure}

As illustrated in Fig. ~\ref{fig:Attenuation}, the tempered attenuation kernel overcomes the singularity and ensures finite stress-strain at the local point. Besides, the behaviour of the tempered kernel in the far nonlocal region is almost identical to that of the conventional kernel for all values of $\alpha$. In the limit as $\alpha\rightarrow0$, the applied stress is felt all over the bar with little attenuation. On the other hand, in the limit as $\alpha\rightarrow1$, the attenuation kernel approaches the Dirac-delta form implying that the stress is completely attenuated in the immediate neighbourhood of the local point.

As seen, the tempered attenuation kernel does not decay monotonically from the local point and therefore, seems contrary to the behaviour expected from an ideal attenuation kernel. However we stress that, since the two peaks arising due to tempering lie in the immediate vicinity of the local point, the behaviour of the material is not affected significantly. Besides, this undesired tempering effect can easily be compensated by giving appropriate weighting to the material constant $\kappa$.
\par
Replacing the conventional power-law attenuation kernel in equation \eqref{eq4} by its tempered version \eqref{eq5}, we then obtain the constitutive relation of the tempered nonlocal elastic bar as,
\begin{equation}\label{eq6}
\sigma\left(x\right)=E_{0}\left[{\epsilon\left(x\right)}+{\frac{2\kappa}{\Gamma\left(1-\alpha\right)}\int\limits _{0}^{\infty}\frac{\left(1-e^{-\lambda y}\right)\epsilon\left(x-y\right)}{y^{\alpha}}dy}\right].
\end{equation}

\section{Spatial dispersion from tempered fractional calculus}\label{sec:4}

\setcounter{section}{4}
\setcounter{equation}{0}

Since in this section we will transform equation \eqref{eq6} to the Fourier domain, it is essential to first introduce the mathematical framework of dispersion modelling. For a unit amplitude, one-dimensional propagating plane wave, the displacement $u$ is given as:
\begin{equation}\label{eq7}
u\left(x,t\right)=e^{i\left(kx-\omega t\right)},
\end{equation}
where $x$ is space and $t$ is time. In a loss-less medium, the phase velocity is $c_{p}=\omega/k$, the group velocity is $c_{g}=d\omega/dk$, and the two are equal. Besides, the wavenumber $k$ is a real quantity, which due to its directional attributes is also called the wave propagation vector. As long as the total energy is conserved during the wave-material interaction, the convention adopted to model wave attenuation is to assume $\omega$ and $k$ as real and complex quantities respectively. However, there are also some examples where the opposite convention of $\omega$ as a complex quantity and $k$ as a real quantity have been made (see, e.g. dissipation of thermoelastic waves in solids \cite{Banerjee1974}). In this work, we assume the wave propagation vector $k$ to be complex such that
\begin{equation}\label{eq8}
k=k_{r}+ik_{i}, \text{ } k_{r}\geq0 \text{ and } k_{i}\geq0,
\end{equation}
where $k_{r}$ is the wave velocity vector and $k_{i}$ is the wave attenuation vector. Here, instead of following the common practice of representing the real and imaginary parts of $k$ by $\beta_{k}$ and $\alpha_{k}$ (see, e.g. \cite{Naesholm2011, Holm2014, Holm2013, NaesHolm2013a, NaesHolm2013b}), we choose to comply with the notations mentioned in \eqref{eq8} and thereby avoid ambiguities with the fractional exponent $\alpha$ of the attenuation kernel. Moreover, the imaginary part $k_i$ of the wavenumber $k$, which we denote as "wave attenuation", should not be confused with the attenuation kernel $g\left(y\right)$. Substituting equation \eqref{eq8} in equation \eqref{eq7}, the displacement of an attenuating wave in a dispersive medium becomes
\begin{equation}\label{eq9}
u\left(x,t\right)=e^{-k_{i}x}e^{i\left(k_{r}x-\omega t\right)},
\end{equation}
with the phase and group velocities given by the real part of $k$ as $c_{p}=\omega/k_{r}$ and $c_{g}=d\omega/dk_{r}$. Both $k_{r}$ and $k_{i}$ are functions of $\omega$ and also related to each other by the Kramers-Kronig relations due to causality \cite{Szabo1995}. The fact that the Kramers-Kronig relations are fundamentally the Hilbert transform pair has two interesting inferences. First, both $\omega$ and $k$ cannot have imaginary values simultaneously. Second, the wave attenuation is always coupled with its velocity.
\par
Now, we introduce the spatio-temporal Fourier transform as
\begin{equation}\label{eq10}
\mathcal{F}\left[u\left(x,t\right)\right]=\hat{u}\left(k,\omega\right)\triangleq\int\limits _{-\infty}^{\infty}\int\limits _{-\infty}^{\infty}u\left(x,t\right)e^{-i\left(kx-\omega t\right)}dx\mbox{ }dt.
\end{equation}
As we are concerned with the dependency of wave-material interactions on the wave propagation vector $k$ alone, we neglect the time domain and only consider the Fourier transform in the spatial domain. Taking the Fourier transform of equation \eqref{eq6}, we get:
\begin{equation}\label{eq11}
\hat{\sigma}\left(k\right)=E_0\left\{ \hat{\epsilon}\left(k\right)+\frac{2\kappa}{\Gamma\left(1-\alpha\right)}\left(I_1-I_2\right)\right\},
\end{equation}
\vskip -5pt \noindent
where
\vskip -12pt
\begin{equation}\label{eq12}
I_1=\int\limits _{0}^{\infty}\mathcal{F}\left[\epsilon\left(x-y\right)\right]y^{-\alpha}dy, \text{ and}
\end{equation}
\vspace*{-5pt} 
\begin{equation}\label{eq13}
I_2=\int\limits _{0}^{\infty}\mathcal{F}\left[\epsilon\left(x-y\right)\right]e^{-\lambda y}y^{-\alpha}dy.
\end{equation}
According to the mass conservation principle, strain is given as
\begin{center}
$\epsilon\left(x,t\right)\triangleq\frac{\partial u}{\partial x}=ik\cdot u\left(x,t\right)$,
\end{center}
and, its Fourier transform is
\begin{equation}\label{eq14}
\hat{\epsilon}\left(k\right)=ik\cdot\hat{u}\left(k\right).
\end{equation}
Next, we consider Newton's second law of conservation of momentum and the equation of motion for the plane wave, from which we respectively have:
\begin{equation}\label{eq15}
\nabla\sigma\left(x,t\right)=\rho\frac{\partial^{2}}{\partial t^{2}}u\left(x,t\right)=-\omega^{2}\rho\text{ }u\left(x,t\right),\ \,  \text{ and}
\end{equation}
\begin{equation}\label{eq16}
\frac{d^{2}u\left(x,t\right)}{dx^{2}}=\frac{1}{c_{0}^{2}}\frac{d^{2}u\left(x,t\right)}{dt^{2}},
\end{equation}
where, $c_{0}$ is the phase velocity of the wave at zero frequency and $\rho$ is the mass density of the material. Assuming zero initial conditions and integrating on both sides of equation \eqref{eq15}, we get
\begin{center}
$\sigma\left(x,t\right)=-\frac{\omega^{2}\rho}{ik}u\left(x,t\right)$.
\end{center}  The corresponding Fourier transform of $\sigma\left(x,t\right)$ is then,
\begin{equation}\label{eq17}
\hat{\sigma}\left(k\right)=-\frac{\omega^{2}\rho}{ik}\hat{u}\left(k\right).
\end{equation}
Substituting equations \eqref{eq17} and \eqref{eq14} in \eqref{eq11}, we obtain:
\begin{equation}\label{eq18}
-\frac{\omega^{2}\rho}{ik}\hat{u}\left(k\right)=E_0\left\{ ik\cdot\hat{u}\left(k\right)+\frac{2\kappa}{\Gamma\left(1-\alpha\right)}\left(I_1-I_2\right)\right\}.
\end{equation}
The translational property of the Fourier transform further gives,
\begin{equation}\label{eq19}
\mathcal{F}\left[\epsilon\left(x-y\right)\right]=e^{-iky} \cdot \hat{\epsilon}\left(k\right).
\end{equation}
Substituting equations \eqref{eq19} and \eqref{eq14} in equations \eqref{eq12} and \eqref{eq13}, we rewrite the expressions of $I_1$ and $I_2$ as:
\begin{equation}\label{eq20}
I_1=ik \cdot \hat{u}\left(k\right)\int\limits _{0}^{\infty}e^{-iky}y^{-\alpha}dy ,\ \, \text{ and}
\end{equation}
\begin{equation}\label{eq21}
I_2=ik \cdot \hat{u}\left(k\right)\int\limits _{0}^{\infty}e^{-iky}e^{-\lambda y}y^{-\alpha}dy.
\end{equation}
Replacing $I_{1}$ and $I_{2}$ in equation \eqref{eq11} with their respective equivalent forms \eqref{eq20} and \eqref{eq21}, and then making use of the relation $c_{0}=\sqrt{E_0/\rho}$ we readily obtain
\begin{equation}\label{eq22}
\frac{\omega^{2}}{k^{2}}=c_{0}^{2}\left\{ 1+\frac{2\kappa}{\Gamma\left(1-\alpha\right)}\left(\int\limits _{0}^{\infty}e^{-iky}y^{-\alpha}dy-\int\limits _{0}^{\infty}e^{-iky}e^{-\lambda y}y^{-\alpha}dy\right)\right\}.
\end{equation}
Next, we proceed to obtain a closed-form solution of the two integrals. It is straightforward to see that the two integrals are modified forms of the Gamma function whose convergence is established using the dominated convergence theorem (see, Chapter 3 in \cite{Meerschaert2011}). Substituting, $x=\left(\lambda-ik\right)y$ in equation \eqref{eq3}, where $\lambda\geq0$ and $k$ is the complex wavenumber, we obtain a useful expression:
\vskip -10pt
\begin{equation}\label{eq23}
\frac{\Gamma\left(z\right)}{\left(\lambda+ik\right)^{z}}=\int\limits _{0}^{\infty}e^{-iky}e^{-\lambda y}y^{z-1}dy.
\end{equation}
The integrals in equation \eqref{eq22} are then evaluated as
\begin{equation}\label{eq24}
\int\limits _{0}^{\infty}e^{-iky}y^{-\alpha}dy=\left(ik\right)^{\alpha-1}\Gamma\left(1-\alpha\right), \ \,  \text{ and}
\end{equation}
\begin{equation}\label{eq25}
\int\limits _{0}^{\infty}e^{-iky}e^{-\lambda y}y^{-\alpha}dy=\left(\lambda+ik\right)^{\alpha-1}\Gamma\left(1-\alpha\right).
\end{equation}
The calculation of similar integrals but for a different exponent of the integrating variable can also be found in a recent paper (see, Theorem 2.1 in \cite{Sabzikar2015}). Now, substituting the value of the integrals from equations \eqref{eq24} and \eqref{eq25} in equation \eqref{eq22}, we finally have
\begin{equation}\label{eq26}
\frac{\omega^{2}}{k^{2}}=c_0^{2}\left[1+2\kappa\left\{ \left(ik\right)^{\alpha-1}-\left(\lambda+ik\right)^{\alpha-1}\right\} \right].
\end{equation}
Equation \eqref{eq26} is the required spatial dispersion relation which governs the propagation of an elastic wave in a tempered nonlocal elastic bar. It is essential to note that the tempering parameter $\lambda$ has a dimension of $\left[L\right]^{-1}$, one the other hand, the material constant $\kappa$ has a variable dimension of $\left[L\right]^{1-\alpha}$.
\par
Further, in the absence of nonlocality i.e., if $\lambda=0$, the spatial dispersion relation \eqref{eq26} reduces to the classical case of loss-less dispersion given as $\omega=c_0k$. The equivalent form of the constitutive relation $\sigma\left(x\right)=E_{0}{\epsilon\left(x\right)}$ is obtained if the same substitution is made in equation \eqref{eq6}.

\section{Numerical solution of the dispersion equation and discussion}\label{sec:5}

\setcounter{section}{5}
\setcounter{equation}{0}

The dispersion equation \eqref{eq26} is a nonlinear equation with complex coefficients and fractional-order exponents of complex wavenumber $k$. As it is difficult to obtain the closed-form expressions for real and imaginary parts of $k$, we resort to root finding algorithms. Besides, the choice of numerical algorithms is limited since most of the available methods are usually applicable for equations containing real coefficients and integer-order exponents (see, Chapter 9 in \cite{Press2007}). We even tried to formulate and numerically implement the classical root-finding algorithms (such as, the bisection method and the Newton-Raphson method) in the complex domain. However, the only algorithm which ensured convergence and achieved the desired level of accuracy was the M\"{u}ller's method.
\par
Based on the generalization of the secant method, M\"{u}ller's method uses quadratic interpolation to approximate the given function and then the root of the function is approximated by the root of the interpolating quadratic (see, Chapter 7 in \cite{Chapra2009}). The algorithm requires three distinct guesses which are used for three functional evaluations to start with, but continues with one function evaluation afterwards. The method does not requires evaluation of derivatives of the function and its rate of convergence is about $1.84$, i.e. nearly quadratic so that the number of correct decimal places almost doubles with each iteration. Further, we have also eliminated the potential round-off errors due to subtractive cancellation by implementing double precision calculation followed by an alternative formulation of the roots. For a quadratic equation of the form $ax^2+bx+c=0$, where $a$, $b$ and $c$ are its coefficients, the formula for the root $x$ which we have implemented in the algorithm is $x_{1,2}=-2c/\left(-b\pm\sqrt{b^{2}-4ac}\right)$, instead of its conventional form $x_{1,2}=\left(-b\pm \sqrt{b^{2}-4ac}\right)/\left(2a\right)$ (see, Chapter 3 in \cite{Chapra2009}).
\par
We set the parameters $\lambda=1$, $c_{0}=1$ and $\kappa=1$ to solve the dispersion equation for three different values of $\alpha$; $\alpha = 0.2, \text{ } 0.5 \text{ and } 0.8$. For each $\alpha$ value, the dispersion equation is solved for a wide range of frequency values, $10^{-4} \leq \omega \leq 10^{7}$. Each frequency decade is uniformly sampled into a minimum of four points, however in situations where significant jumps in the values of the roots are observed, a finer sampling of six to eight points is followed. The values of the numerically obtained roots are accurate up to a minimum of six significant figures. However there was an exception for $\alpha=0.8$, where the algorithm could not converge properly in the frequency range of $10^{-1}$-$10^{0}$ Hz, which further illustrates the numerical difficulty associated with the extraction of complex roots. It should be noted that since the coefficients of wavenumber $k$ in equation \eqref{eq26} are complex, roots do not come in conjugate pairs.
\par

\begin{figure} 
\centering
\includegraphics[scale=0.3]{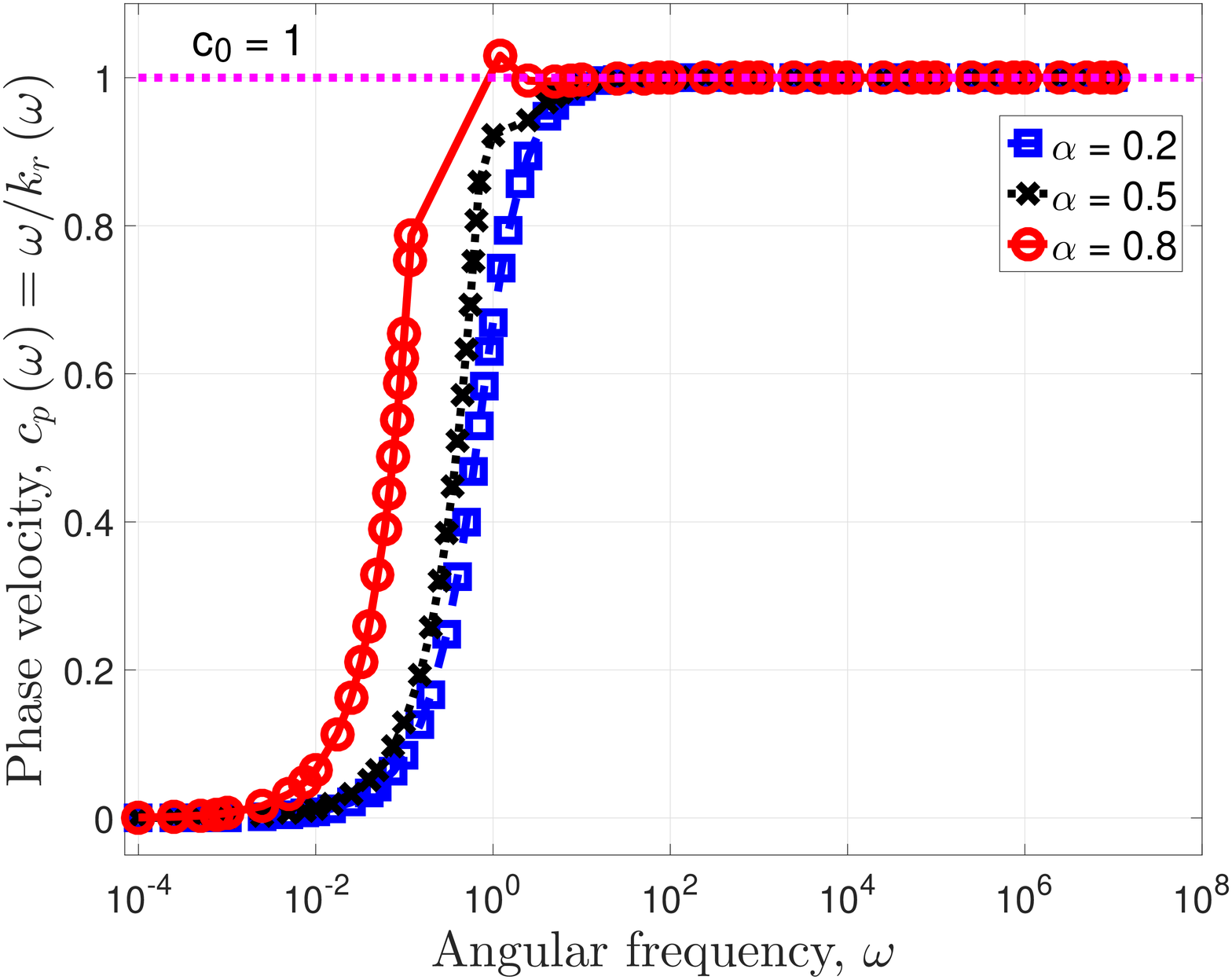}
\includegraphics[scale=0.3]{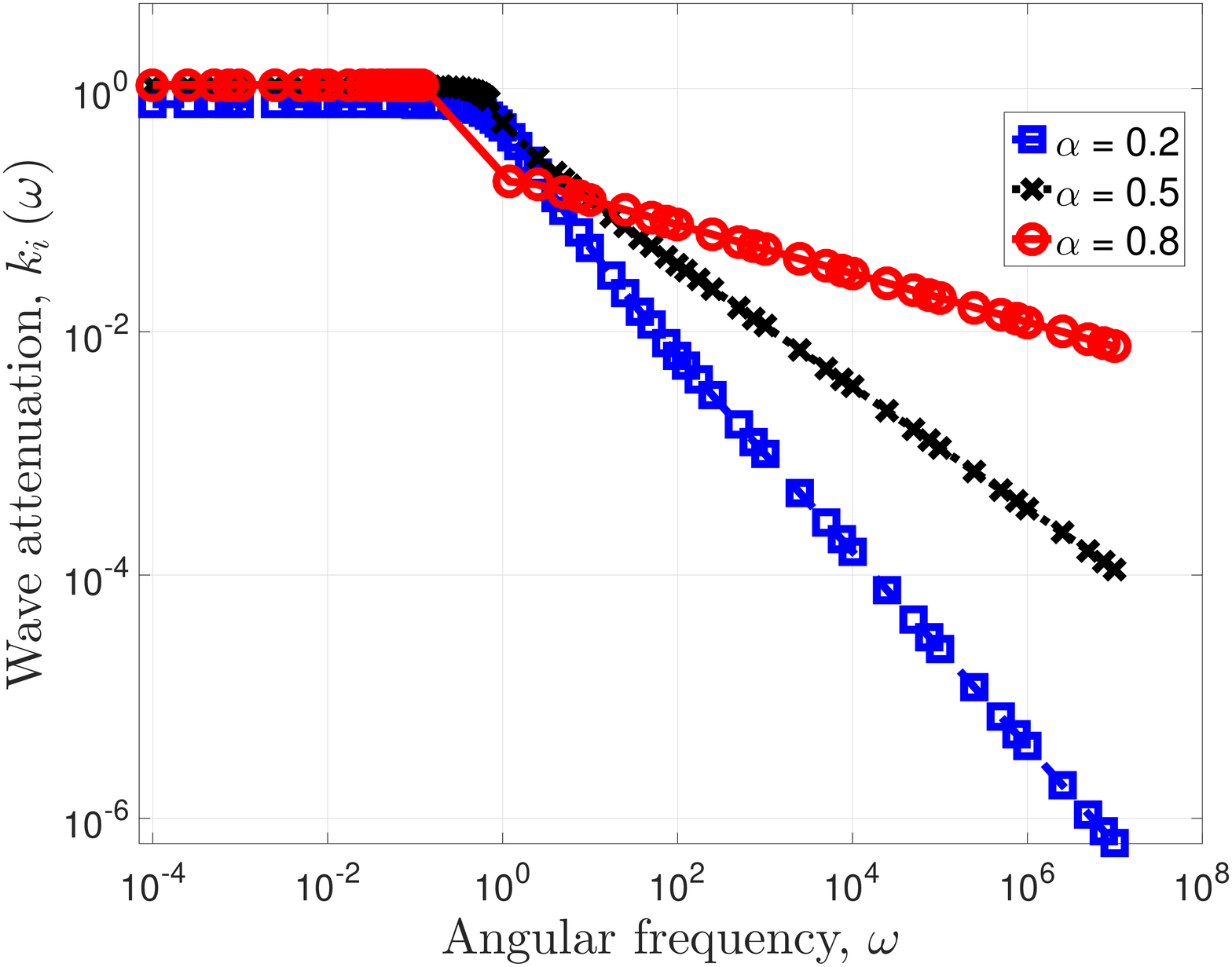}
\smallskip
\caption{\label{fig:Dispersion}Frequency-dependent phase velocity (top plane) and wave attenuation (bottom plane) in a tempered nonlocal elastic bar. The markers; square for $\alpha=0.2$ (blue, dashed curve), cross for $\alpha=0.5$ (black, dotted curve) and circle for $\alpha=0.8$ (red, solid curve), represent the numerically obtained roots. Lack of markers in the frequency decade of $10^{-1}$-$10^{0}$ Hz for $\alpha=0.8$ is due to the non-convergence of M\"{u}ller's algorithm.}
\end{figure}

As illustrated in Fig. ~\ref{fig:Dispersion}, the obtained roots for a given frequency value give the respective phase velocity and wave attenuation. The main observations from the plots can be summarized into four points. First, the pattern of an increased phase velocity with frequency corresponds to anomalous dispersion which is also seen in the case of time-fractional Kelvin-Voigt and Zener models (see, Figures 1 and 2 in \cite{Holm2013}). Second, in the high frequency regime, the phase velocity levels off just like the fractional Zener model. Third, in the low frequency regime of $\omega \leq 10^{-1}$ Hz, wave attenuation is relatively higher than in the intermediate-to-high frequency regime. Fourth, for frequency values $\omega \geq 10^0$ Hz, the straight lines in the log-log attenuation plot, clearly indicate the power-law dependency of wave attenuation $k_i$ on frequency. Further, wave attenuation increases as the value of $\alpha$ is increased.
\par

\par
The obtained dispersion plots can be understood from its underlying physics. Unlike the case of temporal dispersion, spatial dispersion is determined by the wave propagation vector $k$ of the propagating wave. In the case of low frequency waves, the wavelength of the wave is large, as a result the wave interacts with an equivalent larger region of the nonlocal elastic bar. The nonlocal attenuation mechanism which is spread over the length of the bar leads to greater wave attenuation and slows it down. As the frequency of the wave is increased, the size of the wave and hence, the nonlocal region of the bar with which the waves interacts becomes smaller. For very high frequencies, the wave "feels" the material only at a local point and since we have only considered the nonlocal attenuation mechanism, wave attenuation effectively disappears. As explained, a high frequency wave suffers less opposition in the nonlocal elastic bar, and therefore traverses with maximum phase velocity and little attenuation.

\section{Conclusion}\label{sec:6}

\setcounter{section}{6}
\setcounter{equation}{0}

One of the goals which has been achieved in the present work is the understanding of how the phase velocity and wave attenuation are affected as it propagates in a nonlocal elastic material. Even though the wave attenuation in a nonlocal bar appears unusual, it seems very physical. Considering the fact that wave attenuation in a nonlocal bar is negligible in the high frequency regime, such a material if engineered could find applications as an effective channel to transfer energy. We have also established the importance of tempering and the framework of fractional calculus in investigating nonlocal problems where conventional power-law kernels give non-physical results. Further work should include an investigation of the physical implications of the material constant $\kappa$ and the fractional exponent $\alpha$ of the attenuation kernel $g\left(y\right)$.






 \medskip 

 \it

 \noindent
$^{1, 3}$ Dept. of Informatics, 
University of Oslo \\  
P.O. Box 1080, Blindern \\ NO-0316 Oslo, NORWAY \hfill Received: July 15, 2015 \\[3pt]
$^1$ e-mail: vikashp@ifi.uio.no
\hfill Revised: January 28, 2016 \\[5pt]
$^3$ e-mail: sverre@ifi.uio.no \\[10pt]
$^2$ NORSAR\\
P.O. Box 53, N-2027 Kjeller,  NORWAY\\[3pt]
e-mail: peter@norsar.no
\rm

\medskip  

\vskip 0.05cm  
\hrule width40mm height0.10mm 
\vskip 0.05cm

\smallskip

 \noindent
 \smallskip
 \\
 Please cite to this paper as published in:


\textbf{\emph{Fract. Calc. Appl. Anal.}, Vol. 19, No 2 (2016), pp. 498-515,\\
 \textit{\href{DOI: 10.1515/fca-2016-0026}{DOI: 10.1515/fca-2016-0026}}}


\end{document}